\documentclass[11pt,tightenlines,eqsecnum,floats,aps,amsmath,amssymb,nofootinbib,prd,shownopacs]{revtex4}
\usepackage{amsmath,amssymb,amsfonts,amsthm}
\usepackage{graphicx}
\usepackage{enumerate} % advanced enumerate environment
\usepackage{colordvi} % for color text
\usepackage[mathscr]{eucal}
\usepackage[latin1]{inputenc}
\usepackage{amsmath}
\usepackage{amsfonts}
\usepackage{amssymb}
\usepackage{graphicx}

\def\lp{{\ell}_{\rm Pl}}

\def\q{\mathring{q}}

\newcommand{\heff}{{\cal H}_{\mathrm{eff}}}

\newcommand{\p}{\partial}

\newcommand{\f}{\frac}

\usepackage{enumerate}

\def\f{\frac}

\def\d{\textrm{d}}

\usepackage{enumerate}

\newcommand{\be}{\nopagebreak[3]\begin{equation}}
\newcommand{\ee}{\end{equation}}
\newcommand{\ba}{\nopagebreak[3]\begin{eqnarray}}
\newcommand{\ea}{\end{eqnarray}}

\newcommand{\bmult}{\nopagebreak[3]\begin{multline}}
\newcommand{\emult}{\end{multline}}

\def\d{{\rm d}}

\def\lp{{\ell}_{\rm Pl}}

\def\f{\frac}

\def\d{\textrm{d}}

\def\sina{\sin{(\bar{\mu}_1 c_1)}}

\def\sinb{\sin{(\bar{\mu}_2 c_2)}}

\def\sinc{\sin{(\bar{\mu}_3 c_3)}}

\def\cosa{\cos{(\bar{\mu}_1 c_1)}}
\def\cosb{\cos{(\bar{\mu}_2 c_2)}}

\def\Heff{\mathcal{H}_{\rm eff}}
\def\Hmatt{\mathcal{H}_{\rm matt}}

\def\mua{\bar{\mu}_1}
\def\mub{\bar{\mu}_2}
\def\muc{\bar{\mu}_3}
\def\lp{l_{\rm Pl}}
\def\mca{\bar{\mu}_1 c_1}

\def\mcb{\bar{\mu}_2 c_2}

\def\mcc{\bar{\mu}_3 c_3}

\def\d{{\rm d}}
\allowdisplaybreaks

\usepackage{enumerate}
\begin{document}

\title{Curvature invariants, geodesics and the strength of singularities in Bianchi-I loop quantum cosmology}
\author{Parampreet Singh}
\email{psingh@phys.lsu.edu}
\affiliation{Department of Physics and Astronomy, Louisiana State University,
Baton Rouge, Louisiana 70803, USA}

%\pacs{04.60.Pp, 04.60.Kz, 98.80.Qc}

\begin{abstract}
We investigate the effects of the underlying quantum geometry in loop quantum cosmology on spacetime curvature invariants and the extendibility of geodesics in the Bianchi-I model for matter with a vanishing anisotropic stress. Using the effective Hamiltonian approach, we find that even though quantum geometric effects bound the energy density and expansion and shear scalars, divergences of curvature invariants are potentially possible under special conditions. However, as in the isotropic models in LQC, these do not necessarily imply a physical singularity. Analysis of geodesics and strength of such singular events, point towards a general resolution of all known types of strong singularities.
 We illustrate these results for the case of a perfect fluid with an arbitrary finite equation of state $w > -1$, and show that curvature invariants turn out to be bounded, leading to the absence of strong singularities. Unlike classical theory, geodesic evolution does not break down.  We also discuss possible generalizations of sudden singularities which may arise at a non-vanishing volume, causing a divergence in curvature invariants. Such finite volume singularities are shown  to be weak and harmless.
\end{abstract}

%%%%%%%%%%%%%%%%%%%%%%%%
%%%%%%%%%%%%%%%%%%%%%%%%
\maketitle
%%%%%%%%%%%%%%%%%%%%%%%%
%%%%%%%%%%%%%%%%%%%%%%%%
\section{Introduction}

Theorems of Penrose, Hawking and Geroch prove that existence of space-like singularities is a generic feature of Einstein's theory of general relativity (GR) \cite{hawking-ellis}.
These events can be characterized in various ways, such as in terms of divergences of curvature invariants, breakdown of geodesics, tidal forces becoming infinite etc. They signal that the limit of applicability of classical gravity has been reached. It is believed that new physics from a quantum theory of gravity will provide insights on the resolution of singularities, leading to a non-singularity theorem. However, before we implement the techniques of quantum gravity to attempt the resolution of singularities, it is important to distinguish which of these are harmful, and which are harmless. Neither a divergence of curvature invariants implies an end to geodesic extendibility nor geodesic incompleteness necessarily implies existence of a physical singularity. To understand the nature of a singularity one needs to analyze its different properties in unison. In particular, it becomes important to examine the strength of singular events -- whether they lead to an  inevitable destruction of in-falling objects or allow passage of sufficiently strong objects. The former characterize strong singularities, and the latter,  weak singularities \cite{tipler,krolak,clarke-krolak}. It has been conjectured that
strong singularities correspond to events which can not be geodesically extended \cite{tipler2,krolak2}, and thus represent the {\it true} singularities that need to be resolved.

In recent years, methods of loop quantum gravity (LQG) have been successfully used in the quantization of homogeneous spacetimes to address the issue of singularity resolution \cite{mb_rev,as}. In this  canonical framework, known as loop quantum cosmology, a complete quantization of various isotropic spacetimes has been performed for several matter models \cite{aps1,aps2,aps3,bp,warsaw_closed,apsv,acs,kv,szulc,kp,ap}.  These investigations show that in various loop quantized spacetimes, the big bang singularity of GR is replaced by a big bounce which occurs when spacetime curvature approaches Planck regime.\footnote{The existence of bounce has recently been demonstrated in the consistent probabilities framework applied to LQC \cite{craig_singh2}. Similar methods for Wheeler-DeWitt quantization show that the probability for the singularity to occur is unity \cite{craig_singh1}.} These results have been extended to the anisotropic spacetimes where the quantum evolution has been shown to be non-singular \cite{awe2,awe3,we1,bianchi_madrid}, and rich physical implications have been derived \cite{cv,csv,roy_bianchi1,artym_bianchi1,ck,gs1}. The resulting physics of LQC stands in striking contrast to those in Wheeler-DeWitt quantum cosmological models where singularity resolution has remained a formidable challenge to overcome.  A key aspect of LQC which separates it from previous attempts is the underlying quantum geometry inherited from LQG. It has various profound implications. Not only it is responsible for the quantum Hamiltonian constraint to be a quantum difference equation in the geometric representation in LQC, it also leads to upper bounds on energy density \cite{aps3,acs,awe2,cs09,gs1} and expansion and shear scalars of the congruence of geodesics \cite{csv,cs09,ps09,gs1}. Existence of these bounds provide important insights on singularity resolution and also the new physics in loop quantum spacetimes. As an example, a spatially flat isotropic universe sourced with a matter content with equation of state satisfying weak energy condition (WEC) bounces in LQC when the upper bound for energy density is saturated \cite{ps06,aps3}. Thus, avoiding the big bang/crunch singularity.  On the other hand, if it is filled with matter which violates WEC, it recollapses when the maximum of energy density is reached, and avoids big rip singularity\footnote{The big rip singularity in GR occurs when volume diverges in finite time with a divergences in energy density. For an example of a scalar field model leading to such a singularity, see Ref. \cite{ssd}.}  \cite{sst,brip_lqc}.

  An important question is whether the singularity resolution in LQC is a generic phenomena. In particular, whether all strong singularities of the classical theory are resolved in LQC. This question was recently addressed by the author in the spatially flat isotropic model using the effective dynamics of LQC resulting from an effective Hamiltonian \cite{ps09}. The effective Hamiltonian approach, based on the geometrical formulation of quantum mechanics \cite{aa_ts}, provides an invaluable tool to understand the underlying discrete quantum evolution in terms of variables in the continuum effective spacetime.
  The resulting physics obtained from the effective dynamical equations has been extensively tested to confirm with the underlying quantum difference equation for isotropic  and anisotropic models \cite{aps3,apsv,bp,ap,szulc_b1,madrid_bianchi1}. The modified dynamics from the effective Hamiltonian has been used to prove that all strong singularities of the spatially flat isotropic LQC are resolved. These include big bangs/crunches and also finite volume singularities, such as the big freeze singularity \cite{bfs}, which can arise in models with a generalized equation of state.
The effective spacetime of spatially flat isotropic LQC turns out to be geodesically complete \cite{ps09}. Investigations carried out for spatially curved isotropic models in LQC, strongly suggests similar results \cite{sv}. It is interesting to note that these studies also find events where curvature invariants diverge in isotropic LQC. Such events correspond to the sudden singularities, which occur at a finite volume and energy density but have a divergent pressure and have been recently studied in GR \cite{sudden,sudden1,lazkoz,lazkoz2}\footnote{For a discussion of such singularities in modified gravity scenarios, see Ref. \cite{lazkoz3} and reference there in.}. All of the events where curvature invariants diverge in isotropic LQC, turn out to be weak singularities which are harmless.\footnote{The first example of such a `singular' event in LQC was first reported for a scalar field model in Ref. \cite{portsmouth}. It was then shown that such events can arise quite commonly in isotropic LQC with a suitable choice of generalized equation of state \cite{ps09}.}

The goal of this work is to analyze these questions  for Bianchi-I models in LQC using the effective Hamiltonian approach. Due to an interplay of the non-vanishing Weyl curvature with the Ricci curvature, Bianchi-I spacetimes provide one of the simplest expositions of the rich structure of singularities in GR, and thus an opportunity for a non-trivial and important generalization of the results on generic resolution of strong singularities so far proved in isotropic models in LQC. For perfect fluids, depending on the equation of state of matter and the initial anisotropy, singularities can be of the form of a barrel (one of the scale factors is finite and other two vanish), a cigar (one of the scale factors diverges and other two vanish), a pancake (one of the scale factors vanishes and other two approach finite non-zero values) or a point which are isotropic singularities (all scale factors vanish)  \cite{bianchi_doro,bianchi_thorne,ellis_1967,jacobs_1968,ellis_mc1,mccallum}.   In contrast to the isotropic models, where for a fixed equation of state, behavior of energy density is sufficient to capture the details of the spacetime curvature, in Bianchi-I model one must take in to account the important role played by the shear scalar. Roughly speaking, it is both the energy density and the shear scalar which determine whether the expansion scalar behaves more in an isotropic or in an anisotropic way. This is also reflected in the effective dynamics of Bianchi-I model in LQC, where bounces do not occur at maximum value of energy density, but when at least the energy density or the shear scalar approach  values in the Planck regime \cite{csv,gs1}. Due to these rich features, analysis of various aspects of singularity resolution, such as the strength of singularities in Bianchi-I model using Tipler and Kr\'{o}lak's conditions is more subtle than in the isotropic models \cite{clarke-krolak}.  To simplify the analysis, we  restrict ourselves to the study of matter with a vanishing anisotropic stress, and aim to answer following questions related to resolution of singularities in the effective dynamics of Bianchi-I LQC. First, are the curvature invariants bounded in LQC or as in the isotropic spacetimes they can potentially diverge? If so, under what conditions such divergences occur?  Secondly, under what circumstances the geodesic equations in the effective spacetime breakdown? In particular, can geodesics be extended beyond the events where curvature invariants diverge? Finally, do these events correspond to strong or weak singularities?

The analysis carried out in this work and the main results are organized as follows. In Sec. II, for completeness,  we summarize the dynamical equations obtained from the connection-triad variables for the orthogonal Bianchi-I model in GR as well as in effective spacetime description of LQC. Here we demonstrate the way universal bounds on energy density, expansion and shear scalars of geodesic congruences are reached. (For further details on this part, we refer the reader to Refs. \cite{cv,csv,cs09,gs1}). These dynamical equations are used to obtain the expressions for curvature invariants -- Ricci scalar, Kretschmann scalar and the square of the Weyl curvature in Sec III. We then find the conditions under which curvature invariants can potentially diverge in the effective dynamics of LQC. In Sec. IV, we first analyze the behavior of null geodesics and obtain the conditions for which they can break down in the effective spacetime of Bianchi-I model in LQC. These conditions show that events where curvature invariants may diverge in LQC are not necessarily geodesically inextendible. We then perform an analysis of the strength of singular events. Interestingly, the conditions where singularities can potentially be strong turn out to be identical to those which determine the inextendibility of geodesics. In Sec. IVC, we discuss the physical implications of these results for a (non-viscous) perfect fluid with an arbitrary finite equation of state $w > -1$. For this model, we show that unlike the classical theory, there exist no curvature divergent events in LQC. The effective spacetime in LQC turns out to be geodesically complete and devoid of any strong singularities. This is followed by a discussion of the fate of certain types of exotic singularities which may arise with a generalized equation of state. In particular, we analyze the
singularities which may occur at finite volume with a divergence in curvature invariants. Such events, which generalize sudden \cite{sudden} and big freeze singularities \cite{bfs}  of isotropic models to anisotropic spacetimes, are shown be geodesically extendible weak singularities. We summarize the results in Sec. V.

\section{Classical and Effective Hamiltonian in Ashtekar variables: some key features}

 We consider a homogeneous (orthogonal) Bianchi-I  spacetime with a spatial manifold ${\mathbb R^3}$. In order to introduce a symplectic structure, one needs to introduce a fiducial cell ${\cal V}$ on the manifold.\footnote{The cell plays the role of the infra-red regulator which is removed by taking the limit ${\cal V} \rightarrow {\mathbb R}^3$. Physical predictions must not depend on the choice of this cell. This holds for the analysis in this work, as is for several works in LQC. For a discussion on ramifications of these limits and invalidity of unphysical quantizations, see ref. \cite{cs08}.} The cell has a fiducial volume $V_o = l_1 l_2 l_3$, where $l_i$'s denote the coordinate lengths,  with respect to the fiducial metric $\q_{ab}$ on ${\mathbb R^3}$. The physical spacetime metric for the orthogonal Bianchi-I spacetime is given by,
 \be\label{metric}
d s^2 = -N^2 \, d t^2 + a_1^2 \, d x^2 + a_2^2 d y^2 + a_3^2 d z^2
\ee
where $N$ is the lapse function. The basic gravitational variables in LQG are the matrix valued Ashtekar-Barbero connection $A^i_a$ and triad $E^a_i$. Due to underlying symmetries of the spacetime, these are symmetry reduced to connection $c_i$ and triads $p_i$ (where $i = 1,2,3$) \cite{ap_bianchi,chiou,awe2}. The triads are kinematically
related to the directional scale factors as
to the triad components $p_i$ as
 \be\label{triadsf}
|p_1| = l_2 l_3 \, a_2 a_3, ~~~ |p_2| = l_1 l_3 \, a_1 a_3, ~~~ |p_3| = l_2 l_3 \,a_2 a_3 ~.
\ee
The modulus sign arises because of the orientations of the triad. In the following, we choose the orientation to be positive and fiducial lengths $l_i$ to be unity,  without any loss of generality. In the phase space, the connection and the triad variables satisfy:
\be
\{c_i, p_j\} = 8 \pi G \gamma \delta_{ij} ~
\ee
where $\gamma = 0.2375$ is the Barbero-Immirzi parameter in LQG. After imposition of symmetries, the only constraint we need to solve is the Hamiltonian constraint. This is expressed in terms of the holonomies of connection and triads, and quantized to obtain the physical solutions in LQC. In the following, we first discuss the dynamical equations resulting from the Hamiltonian constraint in GR for the Bianchi-I spacetime. This is followed by the analysis with the effective Hamiltonian of LQC in Bianchi-I spacetime.

\subsection{Classical dynamics}

For lapse $N=1$, the classical Hamiltonian constraint in the symmetry reduced  variables $c_i$'s and $p_i$'s can be written as,
\be\label{clH}
{\cal H}_{\mathrm{cl}} = -\f{N}{8 \pi G \gamma^2 V}{(c_1 p_1 \, c_2 p_2 + c_3 p_3 \, c_1 p_1 + c_2 p_2 \, c_3 p_3)} + {\cal H}_{\mathrm{matt}} ~ ,
\ee
where ${\cal H}_{\mathrm{matt}}$ denotes the matter part of the Hamiltonian constraint. Using, ${\cal H}_{\mathrm{cl}}$, dynamical equations can be obtained by solving for the Hamilton's equations for gravitational and matter phase space variables (such as $\phi, p_\phi$, if we consider a scalar field model). In the gravitational sector, these equations are obtained as follows:\footnote{The calculation for matter part follows the same strategy, and is simpler. For a scalar field, the matter phase space variables satisfy $\{\phi,p_\phi\} = 1$. In this case, Hamilton's equations yield Klein-Gordon equation \cite{csv}.}
\be\label{pdot}
\dot p_i = \{p_i, {\cal H}_{\mathrm{cl}}\} = - 8 \pi G \gamma \, \f{\p {\cal H}_{\mathrm{cl}}}{\p c_i}
\ee
and
\be\label{cdot}
\dot c_i = \{c_i, {\cal H}_{\mathrm{cl}}\} =  8 \pi G \gamma \, \f{\p {\cal H}_{\mathrm{cl}}}{\p p_i} ~,
\ee
where a `dot' denotes time derivative with respect to proper time $t$. Using eq.(\ref{pdot}), a relation between connection components and the directional Hubble rates, defined as $H_i = \dot a_i/a_i$, follows
\be\label{ceq}
c_i = \gamma l_i \dot a_i = \gamma H_i a_i~.
\ee
Considering matter with a vanishing anisotropic stress, i.e. when $\rho(p_1,p_2,p_3) = \rho(p_1 p_2 p_3)$,  $(c_i p_i - c_j p_j)$ turn out to be constant of motion \cite{cv,csv}. Using above Hamilton's equations, this in turn implies that,
\be\label{cipi}
c_i p_i - c_j p_j = V(H_i - H_j) = \gamma \kappa_{ij} ~,
\ee
where $\kappa_{ij}$ is a constant anti-symmetric matrix. This feature of GR, that $V (H_i - H_j)$ are constants of motion, results in the characteristic dependence of shear scalar on $1/V^2$. To understand this, let us recall that the covariant derivative of the unit fluid velocity $v^\alpha$ tangent to time-like geodesics, can be written as
\be
v_{\mu;\nu} = \theta_{\mu \nu} + \omega_{\mu \nu}
\ee
where $\theta_{\alpha \beta}$ is the expansion tensor and $\omega_{\alpha \beta}$ is the vorticity tensor.\footnote{The expansion tensor can be introduced in a similar way for null geodesics \cite{poisson}.} Since the considered fluid velocity is orthogonal to the spatial hypersurface, $\omega_{\mu \nu}$ vanishes. The expansion tensor can be expressed in terms of  traceless and trace parts:
\be
\theta_{\mu \nu} = \f{1}{3} (g_{\mu \nu} + v_\mu v^\nu) + \sigma_{\mu \nu} ~.
\ee
where $\theta$ is the expansion scalar:
\be\label{thetadef}
\theta = \dot V/V = (H_1 + H_2 + H_3),
\ee
and $\sigma_{\mu \nu}$ is the shear tensor, satisfying $\sigma_{\mu \nu} v^\nu = 0$. Its magnitude defines the shear scalar $\sigma^2$:
\be\label{sigmadef}
\sigma^2 = \f{1}{3} \left((H_1 - H_2)^2 + (H_2 - H_3)^2 + (H_3 - H_1)^2\right) ~.
\ee
Another measure of anisotropic shear which is used commonly is the scalar $\Sigma^2$, defined as $\Sigma^2 := \sigma^2 V^2/6$. Using
eq. (\ref{cipi}), it turns out to be
\be
\Sigma^2 = \kappa_{12}^2 + \kappa_{23}^2 + \kappa_{31}^2 ~.
\ee
The scalar $\Sigma^2$ is thus a constant of motion in GR for matter with a vanishing anisotropic stress, and thus $\sigma^2 \propto 1/V^2$ in GR.
In the classical theory, the expansion scalar $\theta$ (\ref{thetadef}) and shear scalar $\sigma^2$ are related to the energy density, defined as $\rho = {\cal H}_{\mathrm{matt}}/V$, as
\be\label{fried_cl}
\f{\theta^2}{9} = \f{8 \pi G}{3} \rho + \sigma^2 ~.
\ee
Since, mean Hubble rate $H = \theta/3$,  above equation provides a generalization of the isotropic Friedmann equation in the Bianchi-I spacetime.   Solutions of the dynamical equations have been extensively studied for different matter content \cite{bianchi_doro,bianchi_thorne,ellis_1967,jacobs_1968}. These exhibit singular behavior at vanishing physical volume where geodesics break down \cite{hawking-ellis}.  All the singularities studied so far in the classical Bianchi-I model are strong in nature which cause inevitable destruction of all the detectors falling in them \cite{tipler,krolak}. Late time evolution of the universe, whether it becomes isotropic or remains anisotropic depends on the matter content. From (\ref{fried_cl}), we find that the expansion scalar isotropizes, at large volume, if the universe is filled with matter such as dust or radiation or when ever the energy density of the  matter decays slower than the anisotropic shear scalar $\sigma^2$ in GR.

\subsection{Effective dynamics}
Loop quantization of Bianchi-I spacetimes has been rigorously performed for a massless scalar field and implications for the quantum theory have been studied recently \cite{awe2} (see also Ref. \cite{bianchi_madrid}).\footnote{For earlier works in LQC on Bianchi-I spacetime, see Refs. \cite{chiou,b1_old,szulc_b1,madrid_bianchi1,ck_b1}.} To quantize, the classical Hamiltonian constraint is expressed in terms of the elementary variables for the quantum theory: the holonomies of the connection $A^i_a$ taken over closed loops and the fluxes of the triads $E^a_i$ (which turn out to be proportional to triads). The procedure leads to a non-local nature of the field strength of the connection  in the quantum theory, resulting in  a quantum difference equation in geometric representation. The quantum difference equation turns out to be non-singular and results in a continuous differential equation for Wheeler-DeWitt theory when spacetime curvature becomes small.
 An important technique to extract new physics resulting from loop quantization of cosmological spacetimes is the effective Hamiltonian method. The effective spacetime description of LQC is derived using techniques of the geometric formulation of quantum mechanics \cite{aa_ts}. In this formulation, one treats the space of quantum states as an infinite dimensional quantum phase space and seeks a faithful embedding of the classical phase space in the latter. The underlying procedure, requires a judicious choice of coherent states, and results in an effective Hamiltonian up to well controlled approximations.\footnote{There also exist terms proportional to quantum fluctuations resulting from the quantum properties of the state. These turn out be negligible for the physical universes which grow to a macroscopic size, as is demonstrated by several numerical simulations \cite{aps2,aps3,apsv,bp,ap}. Our analysis will assume that such terms can be neglected for the effective Hamiltonian for Bianchi-I model in LQC.} Using the effective Hamiltonian, modified dynamical equations incorporating quantum geometric effects can be derived \cite{jw,vt,psvt}. Resulting effective equations have been tested in different models using extensive numerical simulations, and have been demonstrated to capture various details of underlying quantum evolution to an excellent accuracy for isotropic models \cite{aps2,aps3,apsv,bp,ap} and also anisotropic models \cite{szulc_b1,madrid_bianchi1}. In the Bianchi-I model, the effective Hamiltonian has been used to understand various physical implications of the loop quantization \cite{cv,csv,cs09,ck,roy_bianchi1,artym_bianchi1} and have also been studied in the context of Gowdy spacetimes \cite{madrid}. As in these works, we would assume that the effective spacetime description for the Bianchi-I model to be valid in our analysis.

For lapse $N=1$, the effective Hamiltonian constraint for the Bianchi-I model in LQC is given by
\be\label{effham}
\heff =  - ~\f{1}{8 \pi G \gamma^2 V}\left(\f{\sin(\bar \mu_1 c_1)}{\bar \mu_1} \f{\sin(\bar \mu_2 c_2)}{\bar \mu_2}  p_1 p_2 + \mathrm{cyclic} ~~ \mathrm{terms}\right) ~~ + ~~ {\cal H}_{\mathrm{matt}}~  \\
\ee
where $\bar \mu_i$ are proportional to the length of the edges of the loop over which holonomies are evaluated and are given by,
%proportional to the square root of the minimum eigenvalue of the area operator in LQG: $\Delta = 4 \sqrt{3} \pi \gamma \lp^2$
\be \label{mub1}
\bar \mu_1 = \lambda \sqrt{\f{ p_1 }{p_2 p_3}}, ~~~ \bar \mu_2 = \lambda \sqrt{\f{p_2}{p_1 p_3}}, ~~~ \mathrm{and} ~\bar \mu_3 = \lambda \sqrt{\f{p_3}{p_1 p_2}} ~ .
\ee
Here $\lambda$ denotes the square root of the minimum eigenvalue of the area operator in LQG: $\Delta = 4 \sqrt{3} \pi \gamma \lp^2$, which is the minimum area to which the loop can be shrunk in the quantum theory \cite{awe1}. The matter part in the effective Hamiltonian is treated as obtained from Fock quantization.\footnote{In principle, there can be quantum geometric modifications to the matter Hamiltonian, originating from the inverse triad terms in the loop quantization. However, these can only be meaningfully defined for compact topologies and are significant only at the scales of Planck length \cite{as,aps3}.}  The effective Hamiltonian leads to the modified dynamical equations, which can be obtained using eqs.(\ref{pdot}) and (\ref{cdot}). An immediate consequence of the equation of motions for the triads is that
the connection components, unlike in the classical theory, are not proportional to $\dot a_i$. Due to this reason, the directional Hubble rate for a particular scale factor is determined by all the connection components. As an example, the relation for $H_1$ is given by
\be \label{H_bianchi1}
H_1  = \frac{1}{2\gamma \lambda}\left( \sin{(\mua c_1-\mub c_2)}+\sin{(\mub c_2+\muc c_3)}+\sin{(\mua c_1-\muc c_3)} \right)
%H_2 & = &\frac{1}{2\gamma \lambda}\left( \cosa(\sinb+\sinc) +\cosc(\sina+\sinb) -\cosb(\sina+\sinb) \right)\\
\ee
(and similarly for $H_2$ and $H_3$). Here we have used
\be
\label{p1dot}\dot{p_i}=\frac{p_i}{\gamma \lambda} \left(\sin(\bar \mu_j c_j) + \sin(\bar \mu_k c_k) \right)\cos(\bar \mu_i c_i)
\ee
and the relation between the directional Hubble rate and the time derivatives of triads:
\be\label{hubbledef}
H_i \, = \,\frac{1}{2} \, \left(\frac{\dot{p_j}}{p_j} + \frac{\dot{p_k}}{p_k} - \frac{\dot{p_i}}{p_i}\right),
\ee
with $i, j$ and $k$ taking different values corresponding to  the anisotropic directions.\\

Though,  $c_i \neq \gamma a_i H_i$  in LQC, using the Hamilton's equation for connection,
\ba\label{dotc1}
\dot c_1 &=& \nonumber \f{1}{2 p_1 \gamma \lambda}\Bigg[c_2 p_2 \cos(\bar \mu_2 c_2) (\sin(\bar \mu_1 c_1) + \sin(\bar \mu_3 c_3)) + ~c_3 p_3 \cos(\bar \mu_3 c_3) (\sin(\bar \mu_1 c_1) + \sin(\bar \mu_2 c_2)) \nonumber \\
&&
 \hskip1.5cm - ~ c_1 p_1 \cos(\bar \mu_1 c_1) (\sin(\bar \mu_2 c_2) + \sin(\bar \mu_3 c_3)) - ~\bar \mu_1 p_2 p_3 \bigg[\sin(\bar \mu_2 c_2)\sin(\bar \mu_3 c_3) \nonumber  \\ && \hskip1.5cm + \sin(\bar \mu_1 c_1)\sin(\bar \mu_2 c_2) + \sin(\bar \mu_3 c_3)\sin(\bar \mu_1 c_1) \bigg]
\Bigg] ~  + ~ 8 \pi G \gamma \, \sqrt{\f{p_2 p_3}{p_1}} \, \left(\f{\rho}{2} + p_1 \f{\partial \rho}{\partial p_1}\right) ,~ \nonumber \\
\ea
and similarly for $c_2$ and $c_3$, it is straightforward to verify that
\be\label{cipi_lqc}
\f{\d}{\d t} (c_i p_i - c_j p_j) = 0 ~.
\ee
Thus, as in GR, $(c_i p_i - c_j p_j)$ are constants of motion in LQC, equal to $\gamma \kappa_{ij}$ (eq.(\ref{cipi})) \cite{cv,csv}. However, $c_i p_i - c_j p_j \neq V (H_i - H_j)$. Thus, shear scalar $\sigma^2$ in LQC is not proportional to $1/V^2$. In other words, if one defines $\Sigma^2_{\mathrm{LQC}} = \sigma^2 V^2$ in LQC,  $\Sigma^2_{\mathrm{LQC}}$ is not a  constant. In the limit when spacetime curvature becomes small, $\Sigma^2_{\mathrm{LQC}}$ approaches the constant value $\Sigma^2$ of the classical theory.\\

Novel properties of the above effective dynamical equations  have been studied in various works \cite{cv,csv,roy_bianchi1,artym_bianchi1,gs1}. Here we summarize the results for the expansion scalar, energy density and the shear scalar, which are relevant for our work. (We refer the reader to Ref. \cite{gs1} for details). We first note that the directional Hubble rates have a universal maxima, given by
\be\label{hib1_max}
H_i^{\mathrm{(max)}} = \f{3}{2 \gamma \lambda} ~.
\ee
Using eq.(\ref{hubbledef}) in the definition of the expansion scalar, we obtain
\be\label{thetab1}
\theta = \f{1}{2\gamma\lambda}\left( \sin{(\mua c_1+\mub c_2)}+\sin{(\mub c_2+\muc c_3)}+\sin{(\mua c_1+\muc c_3)} \right) ~.
\ee
It has a universal maxima given by $\theta_{\mathrm{max}} = 3/(2 \gamma \lambda)$. Thus, unlike in the classical GR, where the directional Hubble rates and the expansion scalar diverges for arbitrary matter in the Bianchi-I spacetime, they are generically bounded in LQC. The boundedness of the directional Hubble rates immediately implies the same for the shear scalar $\sigma^2$ given by eq.(\ref{sigmadef}), which turns out to be,
\ba
\label{shearb1} \sigma^2 &=& \nonumber \frac{1}{3 \gamma^2 \lambda^2 }\Bigg[ (\cosb(\sina+\sinc)- \cosa(\sinb+\sinc))^2  \\ && +~\mbox{~cyclic terms}~\Bigg] ~,
\ea
with a maxima given by $\sigma^2_{~\mathrm{max}} = 10.125/(3 \gamma^2 \lambda^2)$.
%\be
% {\sigma^2}_{\rm max}=\frac{10.125}{3 \gamma^2 \lambda^2} .
%\ee
Recall that the shear scalar in classical GR diverges as $1/V^2$ as a singularity is approached. However, in LQC, its behavior is very different in the Planck regime where it departs from the classical behavior and is generically bounded without any assumption on the matter content. Finally,
the expression for energy density, defined as $\rho = \Hmatt/V$, can be obtained from the vanishing of the Hamiltonian constraint, $\Heff \approx 0$:
\be
\label{rhob1} \rho=\frac{1}{8 \pi G \gamma^2 \lambda^2}\left(\sina \sinb + \mbox{cyclic terms}\right) ~,
\ee
which is also a universally bounded function with a maxima given by
\be\label{rho_max}
\rho_{\rm max} = \f{3}{8 \pi G \gamma^2 \lambda^2}  \approx0.41 \rho_{\rm Pl} ~.
\ee
Thus, in the loop quantization of  Bianchi-I spacetime, the expansion and the shear scalar of the geodesic congruences and the energy density of the matter content turn out to be generically bounded. This is direct consequence of the underlying discreteness of the quantum geometry in LQC, captured by $\lambda$. In the limit, $\lambda \rightarrow 0$, the quantum discreteness disappears and the behavior of the above physical quantities turns out to be in agreement with GR, diverging as the volume goes to zero.

\section{Curvature invariants}

In GR, one of the characteristic feature of singularities is the associated divergence in the curvature invariants.  The key question is whether the quantum geometric effects encoded in the effective dynamics of LQC lead to a bound on the curvature invariants. The answer to this question is not obvious from the properties of energy density, and the expansion and shear scalars, without further assumptions on the nature of the considered matter or its equation of state.  Since the curvature invariants involve the second order time derivatives of the metric, it is possible that even when $\rho$, $\theta$ and $\sigma^2$ are bounded, for a certain choice of the equation of state $w$, curvature invariants may blow up in the evolution. Examples of such events, exist in isotropic LQC, where they originate from the divergences in the pressure $P$ of the matter content \cite{ps09,sv}.  In the following, we study the properties of the Ricci scalar $R$, Kretschmann scalar $K$ where $K = R_{\alpha \beta \mu \nu} R^{\alpha \beta \mu \nu}$ and the square of the Weyl curvature $C_{\alpha \beta \mu \nu} C^{\alpha \beta \mu \nu}$.\footnote{Other curvature invariants can be studied similarly, or there expressions can be obtained from these invariants. As an example,  properties of another useful invariant $R_{\alpha\beta} R^{\alpha\beta}$ can be obtained using the identity: $R_{\alpha\beta} R^{\alpha\beta} = \tfrac{1}{2} (K - C_{\alpha \beta \mu \nu} C^{\alpha \beta \mu \nu} + R^2/3)$.}

In terms of the directional Hubble rates $H_i$ and the second order time derivatives of the directional scale factors, the expressions for the Ricci scalar, Kretschmann scalar and the square of the Weyl curvature can be written as:
\be\label{ricci}
R = 2 \left(H_1 H_2 + H_2 H_3 + H_3 H_1 + \sum_{i=1}^3 \f{\ddot a_i}{a_i} \right) ~,
\ee

\be\label{kretschmann}
K = 4 \left(H_1^2 H_2^2 + H_1^2 H_3^2 + H_2^2 H_3^2 + \sum_{i=1}^3 \f{\ddot a_i^2}{a_i} \right) ~,
\ee

and
\ba\label{weyl1}
C_{\alpha \beta \mu \nu} C^{\alpha \beta \mu \nu} &=& \nonumber \f{4}{3} \Bigg[H_1^2 H_2^2 + H_2^2 H_3^2 + H_3^2 H_1^2 - H_1 H_2 H_3 (H_1 + H_2 + H_3)  \\
&& \nonumber + ~ \f{\ddot a_1}{a_1} \left(\f{\ddot a_1}{a_1} - H_1 H_2 - H_3 H_1 + 2 H_2 H_3 \right) + ~\f{\ddot a_2}{a_2} \left(\f{\ddot a_2}{a_2} - H_1 H_2 - H_2 H_3 + 2 H_3 H_1 \right) \\
&& \nonumber ~\f{\ddot a_3}{a_3} \left(\f{\ddot a_3}{a_3} - H_2 H_3 - H_3 H_1 + 2 H_1 H_2 \right) - \f{\ddot a_1}{a_1} - \f{\ddot a_2}{a_2} - \f{\ddot a_3}{a_3}\Bigg] ~.
\ea
From  above expressions, it is clear that since the directional Hubble rates in the effective spacetime of Bianchi-I model in  LQC are bounded, as evident from eq.(\ref{H_bianchi1}), the boundedness of curvature invariants is determined by the dynamical equations $\ddot a_i/a_i$.  These can be obtained by using the Hamilton's equations for $p_i$ (eq.(\ref{p1dot})) and $c_i$ (eq.(\ref{dotc1})). As an example, the equation for $\ddot a_1/a_1$ can be written as,
\ba
\f{\ddot a_1}{a_1} &=& \nonumber \f{1}{2 \gamma \lambda} \Bigg[\cos(\mca - \mcb)\, \f{\d}{\d t} (\mca - \mcb) ~ + ~\cos(\mca - \mcc)\, \f{\d}{\d t} (\mca - \mcc) \\&&  \nonumber +  \cos(\mcb + \mcc) \, \f{\d}{\d t} (\mcb + \mcc) + \f{1}{2 \gamma \lambda} \left(\sin(\mca - \mcb) + \sin(\mca - \mcc) + \sin(\mcb + \mcc)\right)^2 \Bigg]\\
\ea
which for matter with a vanishing anisotropic stress, can be expressed in the following form:
\ba\label{ddota1}
\f{\ddot a_1}{a_1} &=& \nonumber - 4 \pi G (\rho + P) + \f{1}{4 V} \left(6 (\cos(\mca - \mcb) \kappa_{21} + \cos(\mca - \mcc) \kappa_{31}) H + (\kappa_{12} + \kappa_{13}) \f{\dot p_1}{p_1} \right. \\&&  \left. + \kappa_{23} \left(\f{\dot p_2}{p_2} - \f{\dot{p_3}}{p_3}\right)\right) + \f{1}{4 \gamma^2 \lambda^2} \, \chi^2 ~,
\ea
where pressure $P = - \partial {\cal H}_{\mathrm{matt}}/\partial V$, and $\chi^2$ is a bounded function defined as
\be
\chi^2 := \left(\sin(\mca - \mcb) + \sin(\mca - \mcc) + \sin(\mcb + \mcc)\right)^2 ~.
\ee
In order to obtain (\ref{ddota1}), we have used the relation between the connection and triad components and the constants $\kappa_{ij}$ (given by eq.(\ref{cipi})). Similar calculations, result in the equations for $\ddot a_2/a_2$ and $\ddot a_3/a_3$. Using these dynamical equations, along with those for the directional Hubble rates (\ref{hubbledef}), expressions for curvature scalars can be obtained by  straightforward calculations.

Substituting eqs.(\ref{H_bianchi1}) and (\ref{ddota1}), and similar equations for the time derivatives of $a_2$ and $a_3$ in  the expression for the Ricci scalar (\ref{ricci}), for matter with a vanishing anisotropic stress, we get,
\ba\label{r_ci}
R &=& \nonumber -24 \pi G (\rho + P) + \f{1}{V}\left(\f{\dot p_1}{p_1} (\kappa_{12} + \kappa_{13}) + \f{\dot p_2}{p_2} (\kappa_{23} + \kappa_{21}) + \f{\dot p_3}{p_3} (\kappa_{31} + \kappa_{32}) \right)\\
&& \nonumber + \f{1}{2 \gamma^2 \lambda^2} \Bigg[ 3 + \cos^2(\bar \mu_1 c_1) \left(\sin^2(\bar \mu_3 c_3) + 4 \sin(\bar \mu_2 c_2) \sin(\bar \mu_3 c_3) - \cos(2 \bar \mu_2 c_2)\right) \\
&& \nonumber ~ + \cos^2(\bar \mu_2 c_2) \left(\sin^2(\bar \mu_1 c_1) + 4 \sin(\bar \mu_1 c_1) \sin(\bar \mu_3 c_3) - \cos(2 \bar \mu_3 c_3)\right) \\
&& \nonumber ~ + \cos^2(\bar \mu_3 c_3) \left(\sin^2(\bar \mu_2 c_2) + 4 \sin(\bar \mu_1 c_1) \sin(\bar \mu_2 c_2) - \cos(2 \bar \mu_1 c_1)\right) \\
&&  - \left(\sin^2(\bar \mu_1 c_1) \sin^2(\bar \mu_2 c_2) + \sin^2(\bar \mu_1 c_1) \sin^2(\bar \mu_3 c_3) + \sin^2(\bar \mu_2 c_2) \sin^2(\bar \mu_3 c_3)\right) \Bigg] ~.
\ea
Let us analyze the behavior of Ricci scalar in the effective spacetime. From the dynamical equations in Sec.II (eqs.(\ref{p1dot}) and (\ref{rhob1})),  $\rho$ and $\dot p_i/p_i$ are universally bounded in LQC. Further, $\kappa_{ij}$ are the constants equal to $(c_i p_i - c_j p_j)$ in LQC (eq.(\ref{cipi_lqc})). These imply, that the Ricci scalar can diverge only if volume vanishes and/or the pressure diverges at a finite energy density.  Note that if we consider the case of a fluid with a finite equation of state and the dynamical evolution is such that the volume does not vanish, then Ricci scalar does not diverge. In such a case, Ricci scalar is always bounded. An example of such a matter content is the case of perfect fluids  which will be discussed in Sec. IVC.   \\

  The computation of the Kretschmann scalar for matter with a vanishing anisotropic stress can be performed in an analogous way. Using the equations for $H_i$ and $\ddot a_i/a_i$, in eq.(\ref{kretschmann}), we obtain
\ba\label{k_ci}
K &=& \nonumber \f{\chi^2}{2 \gamma^4 \lambda^4} \left\{(2 + \cos(\mca-\mcb-2\mcc) + \cos(\mca-\mcb+2\mcc)) \sin^2\left(\tfrac{\mca+\mcb}{2}\right)\right. \\
&& \nonumber \left. ~~~ + \sin^2(\mca-\mcb) + 2 (\cos(\mcb) - \cos(\mca)) \sin(\mca - \mcc) \sin(\mcc) \right\}\\
%&& \nonumber ~~~ \times \left(\sin(\mca - \mcb) + \sin(\mca - \mcc) + \sin(\mcb + \mcc)\right)^2 \\
&& \nonumber + ~ \f{1}{4 \gamma^2 \lambda^2 V^2} \Bigg[\f{\dot p_1}{p_1} (\kappa_{12} + \kappa_{13}) + \kappa_{23} \left(\f{\dot p_2}{p_2} - \f{\dot p_3}{p_3}\right) + 6 H (\kappa_{21} \cos(\mca-\mcb) + \kappa_{31} \cos(\mca - \mcc)) \\
&&\nonumber  ~~~ - 16 \pi G (\rho + P) V\Bigg] \Bigg[\gamma \lambda \left\{6 H \gamma \lambda(\kappa_{21} \cos(\mca-\mcb) + \kappa_{31} \cos(\mca - \mcc)) \right. \\
&& \nonumber ~~~~~ + \kappa_{23} (\cos(\mcb)(\sin(\mca) + \sin(\mcc)) - \cos(\mcc) (\sin(\mca) + \sin(\mcb))) \\
&&  \nonumber ~~~~~+ \left. (\kappa_{12} + \kappa_{13}) \cos(\mca) (\sin(\mcb) + \sin(\mcc))\right\} + 2 V \left(- 8 \pi G \gamma^2 \lambda^2 (\rho + P) + \chi^2\right)\Bigg] \\
&& + ~~ \mathrm{cyclic} ~ \mathrm{terms} ~.
\ea
Analysis of various terms in the above expression shows that as in the case of the Ricci scalar, the Kretschmann scalar can also diverge in LQC. Potential divergent terms are proportional to $1/V^2$, $P/V$ and terms linear and quadratic in pressure. If the dynamical evolution restricts these possibilities, then $K$ is bounded.

Similarly, we can compute the square of the Weyl curvature. In order to obtain its expression, it is useful to note that eq.(\ref{weyl1}) can be written in the following symmetric form
\ba
C_{\alpha \beta \mu \nu} C^{\alpha \beta \mu \nu} &=& \nonumber \f{4}{3} \bigg[\dot H_1(\dot H_1 - \dot H_2) - \dot H_1 ((H_2 - H_3)^2  + H_1 H_2 + H_2 H_3 - 2 H_1^2) \\
&& ~~~ + H_1^2(H_1^2 + H_2 H_3 - H_1 H_2 - H_3 H_1) ~ + ~ \mathrm{cyclic} ~ \mathrm{terms} \bigg] ~.
\ea
Using the equations for the directional Hubble rates and its time derivative (obtained using (\ref{ddota1})), in the above equation, we obtain the following expression for a matter with a vanishing anisotropic stress,
%\ba\label{w_ci}
\begin{align}\label{w_ci}
C_{\alpha \beta \mu \nu} C^{\alpha \beta \mu \nu} =& \nonumber - \f{1}{32 \gamma^2 \lambda^2 V} \bigg[ 2 \cos(2(\mca - \mcb) + \cos(\mca + \mcb) - \cos(2(\mca + \mcb)))\\
&  \nonumber  ~~~ + 5 \cos(\mca - \mcb) + \cos(2 \mcb) (3 - 2 \cos(\mcc)) + \cos(2 \mca)(- 6 + \cos(2 \mcc)) \\
&  \nonumber  ~~~ + 3 \cos(2 \mcc) + 3(\sin(2 \mca) - 4 \cos(\mcb)(\sin(\mca) + \sin(\mcb))) \sin(2 \mcc) \\
&  \nonumber ~~~ - 2 \left\{ 8 \cos^2(\mca) \sin(\mcb) \sin(\mcc) \right. + 2 \sin(\mca) \left[-\cos(2\mcc) \sin(\mcb) \right. \\
&  \nonumber \left. ~~+ 6 \cos(\mcb) \cos(\mcc) (\sin(\mca) + \sin(\mcb)) - 2 \cos^2(\mcb) \sin(\mcc) \right] \\
&  \nonumber ~~~ + 3 \cos(\mca) \left(-\cos(\mcc) + \cos(2 \mcb + \mcc) + \cos(\mcb + 2\mcc)\right. \\
&  \nonumber \left. ~~ - 2 \sin(\mca) \sin(\mcb + \mcc)\right)\left. \right\} \bigg] \times \bigg[ \f{\dot p_1}{p_1} (\kappa_{12} + \kappa_{13}) + \kappa_{23} \left(\f{\dot p_2}{p_2}- \f{\dot p_3}{p_3}\right)\\
&  \nonumber   ~~~ + 6 H (\kappa_{21} \cos(\mca-\mcb) + \kappa_{31} \cos(\mca - \mcc)) - 16 \pi G (\rho + P) V\bigg]\\
&\nonumber + ~ \f{\chi^2}{4 \gamma^4 \lambda^4} \left(\cos(\mcc)(\sin(\mca) + \sin(\mcb)) - \cos(\mca)(\sin(\mcb) + \sin(\mcc))\right) \\
& \nonumber  ~~~ \left(\cos(\mcb)(\sin(\mca) + \sin(\mcc)) - \cos(\mca)(\sin(\mcb) + \sin(\mcc))\right)  \\
& \nonumber -  \f{1}{16 V^2} \bigg[ \f{\dot p_1}{p_1} (\kappa_{31} - \kappa_{12}) + \kappa_{23} \left(\f{\dot p_3}{p_3}- \f{\dot p_2}{p_2}\right) + 16 \pi G (\rho + P) V\\
&  \nonumber   ~~~ + 6 H (\kappa_{12} \cos(\mca-\mcb) - \kappa_{31} \cos(\mca - \mcc))\bigg] \times \bigg[\kappa_{12} \left(\f{\dot p_1}{p_1} + \f{\dot p_2}{p_2} \right) - \f{\dot p_3}{p_3} (\kappa_{31} + \kappa_{23}) \\
&  \nonumber ~~~ - 6 H \left(2 \kappa_{12} \cos(\mca - \mcb) - \kappa_{31} \cos(\mca - \mcc) - \kappa_{23} \cos(\mcb - \mcc) \right) \bigg] \\
& + ~ {\mathrm{cyclic}} ~ {\mathrm{terms}} ~.
% && + ~ {}^{(II)}C_{abcd}C^{abcd} ~ + ~  {}^{(III)}C_{abcd}C^{abcd} ~
%\ea
\end{align}
Though more complicated than the expressions for $R$ and $K$, above expression reveals similar properties regarding the boundedness and possible divergences. As for Ricci and Kretschmann scalars, the square of the Weyl curvature can also diverge in LQC if the pressure becomes infinite and/or volume vanishes at a finite value of the energy density. \\

Let us now summarize the main result of this section. Restricting to the case of matter with a vanishing anisotropic stress, analysis of the behavior of various curvature invariants reveals some important features of loop quantum dynamics in Bianchi-I spacetimes. We have found that in LQC,
even though energy density and the expansion and shear scalars are universally bounded by values determined by eqs.(\ref{thetab1},\ref{shearb1}) and (\ref{rhob1}), curvature invariants can potentially diverge. It is to be noted that for all matter models, which lead to dynamics such that the pressure is bounded and physical volume never vanishes during the evolution, curvature invariants are bounded. However, {\it if there exists} a physical solution such that at a finite value of $\rho, \theta$ and $\sigma^2$,  pressure diverges and/or volume  vanishes {\it then} curvature scalars can diverge. For conventional matter models, such as dust, radiation or stiff matter, these conditions are not satisfied. {\it Conventional singularities for Bianchi-I spacetimes for fluids with a finite equation of state, occur with an associated divergence of $\rho$, $\theta$ and $\sigma^2$. Since these quantities are generically bounded in LQC, all such singularities are avoided in effective spacetime.} In all these cases, curvature invariants turn out to be bounded.  Thus, in contrast to  GR,  conditions for which curvature invariants can diverge in the effective spacetime description of LQC are highly restrictive. This is illustrated in Sec. IVC, where we show that for perfect fluids with $w > -1$, there are no divergences in curvature invariants. On the other hand, in GR, such fluids generically lead to divergences in curvature invariant.

Our result on the curvature invariants  generalize the one for the isotropic models in LQC where it was demonstrated that curvature scalars can diverge if the pressure becomes infinite at a finite value of energy density \cite{ps09}.
We find  that such singularities can also arise in effective spacetime description of Bianchi-I models.
 Possible divergence in curvature invariants leads to some important questions: What is the nature of the singularities associated with events where curvature invariants can potentially diverge in Bianchi-I spacetime in LQC? Certainly, these singularities are rather special as they occur when $\rho, \theta$ and $\sigma^2$ are all finite. However, are these singularities strong or weak? And do they imply, breakdown of geodesics?  We answer these questions in the next section.

\section{Geodesics and strength of singularities: general characteristics and examples}

In order to understand the nature of events where curvature invariants diverge, one must analyze the behavior of geodesics and the strength of such singular events. It is important to note that divergence of curvature invariants though an indication, is not a sufficient condition for a physical singularity to exist. As discussed earlier, a singular event in terms of curvature invariants may allow a safe passage of
detectors, in which case the singularity turns out to be weak. In order to determine whether a singularity is strong or weak, we consider the criterion developed by Tipler and Kr\'{o}lak, which involves integrals of Ricci and Weyl curvature components over null or time-like geodesics \cite{clarke-krolak}. Here, without any loss of generality, we will analyze null geodesics.\footnote{Similar analysis can be carried out for time-like geodesics, and similar conclusions as reached in our analysis can be obtained.} In the following we first analyze the geodesic equations and contrast their behavior in GR and in the effective spacetime description of LQC. We then analyze the conditions for the strength of the singularities. Together with geodesic equations, these conditions enable us to understand general features of singularity resolution in the Bianchi-I spacetime in LQC which point towards the lack of strong singularities. We then consider the case of perfect fluids with a vanishing anisotropic stress and a finite equation of state $w = P/\rho > -1$. Our results show that unlike in GR where such a fluid leads to strong singularities of barrel, cigar, pancake and point types, all strong singularities are absent in LQC and the effective spacetime is geodesically complete. We also discuss the potential possibilities of finite volume singularities which can lead to divergences in curvature invariants. These are possible generalizations of sudden singularities in Bianchi-I models \cite{sudden,sudden1}. We find that such singularities are weak in nature, and geodesics can be extended beyond them.

\subsection{Geodesic evolution in Bianchi-I spacetime}
 For the metric (\ref{metric}), the null geodesics obey,
\be
(u^\alpha)' \, + \, \Gamma^\alpha_{\beta \delta} \, u^\beta u^\delta = 0 ~,
\ee
where the 4-velocities $u^\alpha$: $u^\alpha = \d x^\alpha/\d \tau$, satisfy $u_\alpha u^\alpha = 0$, and a `prime' denotes derivative with respect to the affine parameter $\tau$. %, and $.
Computing the Christoffel symbols for the metric (\ref{metric}), a straightforward calculation leads to the following equations:

\be \label{geod1}
x^{\prime \prime} = -2 x' t' H_1, ~~ y^{\prime \prime} = -2 y' t' H_2, ~~ z^{\prime \prime} = - 2 z' t' H_3, ~~ t^{\prime \prime} = - a_1^2 H_1 \,x'^2 - a_2^2 H_2 \,y'^2 - a_3^2 H_3 \,z'^2
\ee
%where $x' = u^x$, $y' = u^y$, $z' = u^z$ and $t' = u^t$,
and
\be \label{geod2}
x' = \f{k_x}{a_1^2}, ~~~  y' = \f{k_y}{a_2^2}, ~~~ z' = \f{k_z}{a_3^2}, ~~~ t' = \left(\f{k_x^2}{a_1^2} + \f{k_y^2}{a_2^2} + \f{k_z^2}{a_3^2}\right)^{1/2} ~,
\ee
where $k_i$ are constants.
From these equations, one can analyze the geodesic extendibility. In particular, one finds that the geodesic evolution breaks down at a finite value of the affine parameter when any of the directional scale factors $a_i \rightarrow 0$  and/or  the directional Hubble rates $H_i$ become infinite. \\% \rightarrow \pm\infty$ at a finite value of the affine parameter.

To analyze the fate of geodesics, we first consider the cases which are expected to arise in most physical situations. These correspond to the scenarios where in classical evolution,  energy density, expansion and shear scalars diverge as the singularity is approached. From eqs.(\ref{ricci}-\ref{weyl1}), we see that such divergences cause curvature invariants to blow up. These cases have been extensively studied for the perfect fluids \cite{bianchi_doro,bianchi_thorne,ellis_1967,jacobs_1968,ellis_mc1,mccallum}. In terms of the directional scale factors (and their permutations), the resulting singularities are of the following types: (i) barrel, characterized by $a_1 \rightarrow$ finite value, $a_2, a_3 \rightarrow 0$, (ii) cigar, where $a_1 \rightarrow \infty$, $a_2, a_3 \rightarrow 0$, (iii) pancake, where $a_1 \rightarrow 0$, $a_2, a_3 \rightarrow$ finite value, and (iv) point or isotropic, characterized by all scale factors vanishing. For all these singularities, at least one of the scale factors and the physical volume vanishes at the singularity. Since the directional Hubble rates diverge at these singularities in the classical theory, eqs.(\ref{thetadef}, \ref{sigmadef}, \ref{fried_cl}) and eqs.(\ref{ricci}-\ref{weyl1}), imply the divergence of
$\rho$, $\theta$, $\sigma^2$ and the curvature invariants in the classical theory. From eqs.(\ref{geod1}) and (\ref{geod2}), we find that the geodesic equations break down at above singularities in the classical theory.

On the other hand, in LQC, for all such cases geodesic equations do not break down. Classical singularities which occur such that $a_i \rightarrow 0$ and $|H_i| \rightarrow \infty$ simultaneously, are forbidden in LQC. Since $H_i$ is bounded (eq.(\ref{hib1_max})), the scale factors at which above singularities occur are excluded from the effective spacetime. To illustrate, in a typical evolution where a resolution of above singularities occurs, one starts with an initial data which for the classical evolution leads to a vanishing of a scale factor and a divergence of the corresponding directional Hubble rate in a finite value of affine parameter. For concreteness, let these be $a_1$ and $H_1$ respectively, where we assume $H_1$ to be positive and increasing in the backward evolution towards the classical singularity. Let us first note, that
when the spacetime curvature is small ($\bar \mu_i c_i \ll 1$), effective dynamical equations approximate the classical dynamical equations. This can be explicitly seen by comparing the effective Hamiltonian constraint (\ref{effham}) with the classical Hamiltonian constraint (\ref{clH}). In this regime, $H_1$ in LQC approximates its counterpart in GR. However, as $H_1$ becomes large, spacetime curvature increases, and deviations between LQC and GR become significant. Analysis of eq.(\ref{H_bianchi1}) and (\ref{ddota1}) shows that depending on the values of $\bar \mu_i c_i$ in the evolution, $H_1$ attains a maximum value in the Planck regime, and starts decreasing, subsequently vanishing before becoming negative.\footnote{For a pictorial illustration of this behavior in a model in effective dynamics in Bianchi-I spacetime, see Ref. \cite{csv} (Fig. 4).} This causes a turn around of the scale factor $a_1$ in effective dynamics and the singularity is avoided.\footnote{The situation is analogous to isotropic models in LQC, where in the backward evolution towards the big bang, a bounce of the scale factor occurs after the Hubble rate attains its maximum value. For examples, see Ref. \cite{ps06,svv}.} Unlike GR, where $H_1$ diverges  and $a_1$ approaches zero, effective dynamical equations cause a turn around of $a_1$ and the vanishing of $a_1$ does not occur. Similar conclusion holds, if more than one scale factors vanish at a classical singularity with associated divergences in the directional Hubble rates. Thus, due to quantum gravitational modifications encoded in effective dynamics, classical singularities accompanied by $a_i \rightarrow 0$ and a divergence in $H_i$   are avoided. Eqs.(\ref{geod1}) and ({\ref{geod2}), then immediately imply that geodesic evolution in these cases does not break down in the effective spacetime description of Bianchi-I spacetime in LQC.\\

\noindent
{\bf Remark:}  In GR, divergence of $\rho$, $H_i$, $\theta$ and $\sigma^2$ may not always occur at a vanishing physical volume. In isotropic models, singularities where $\rho$ and isotropic Hubble rate diverge at non-vanishing volume have been investigated  recently (see for eg. \cite{ssd,bfs}.
These are the big rip (occurring at infinite volume) and big freeze (occurring at a finite non-zero volume) singularities, which are geodesically inextendible events in GR. In isotropic spacetimes in LQC, such singularities have been shown to be avoided due to the boundedness of Hubble rate in LQC \cite{sst,brip_lqc,ps09,sv}. In the effective dynamics, Hubble rate vanishes in the Planck regime and causes a turn-around of the scale factor. This leads to a recollapse of the universe before scale factors at which above classical singularities occur.
Assuming that such  singularities exist in Bianchi-I models in GR,  these will result in break down of geodesics in the classical theory due to divergences in the directional Hubble rates. Since $H_i$ are bounded in LQC, above singularities will be avoided in LQC in the same way as singularities at vanishing scale factors (with associated divergence in $H_i$) are avoided. From eqs. (\ref{geod1}) and (\ref{geod2}), we find that the geodesic equations will be well behaved for these cases in LQC. \\%  \\

So far we have discussed the cases of singularities where divergence in curvature scalars is accompanied by a divergence in $\rho$, $\theta$ and $\sigma^2$. For all such cases, we have found that geodesic evolution does not break down in effective spacetime description of LQC. We now turn to the cases where curvature invariants can diverge at a finite value of $\rho, \theta$ and $\sigma^2$. Examples of such singularities are not known in Bianchi-I models.\footnote{In the anisotropic models in classical theory, such singularities occurring at finite scale factors have been investigated in Bianchi-$\rm{VII_0}$ spacetime \cite{sudden1}.} Here we would assume that such  possibilities can arise in the physical evolution in LQC. From the previous section, we recall that curvature invariants in LQC can diverge under following conditions:   if the physical volume of the spacetime approaches zero, and/or the pressure becomes infinite in magnitude. {\it If} these conditions are satisfied for physical solutions of effective dynamics in LQC {\it then} following two contrasting scenarios can arise:\\

(i) If in the physical evolution, curvature invariants diverge at a finite value of $\rho$, $\theta$ and $\sigma^2$, and a vanishing volume, (which implies that at least one of the scale factors $a_i$ vanishes) at a finite value of the affine parameter, then geodesic equations (\ref{geod1}) and (\ref{geod2}) break down. If such  curvature invariant diverging events exist, then geodesic evolution is not complete.\footnote{However, there may still be no physical singularity. An example being if the spacetime is maximally extendible.}\\

%% geodesic extendibility breaks down only if such a divergence occurs at a finite value of affine parameter. Since H_i are bounded %it is possible to write scale factor as an integal: $a \sim \int e^{f t}$ where $f$ is a bounded function. One may argue that scale %factor goes to zero only in infinite time. However, the same happens for deSitter case, where the scale factor vanishes in a finite %value of affine parameter and geodesics break down. The same can happen here for some choice of H.

(ii) If in the physical evolution, pressure becomes infinite at a finite value of $\rho$, $\theta$ and $\sigma^2$, and a non-vanishing values of the scale factors, and hence non-vanishing volume, then the geodesic equations are well behaved. For these curvature invariant diverging events, geodesic evolution is complete in the effective spacetime description of LQC.  The behavior in this case is similar to the case of pressure singularities occurring in isotropic models in LQC \cite{portsmouth,ps09,sv}, where though for an exotic choice of matter, curvature invariants diverge in the evolution, but the spacetime turns out to be geodesically complete \cite{ps09}.\footnote{Geodesics can be extended beyond such events even in GR \cite{lazkoz}.}\\

To summarize this part, we find that irrespective of the choice of matter, geodesic evolution in the effective spacetime description of Bianchi-I LQC is complete for all the cases except the isolated case when physical evolution may allow divergence of curvature invariants at finite values of $\rho$, $\theta$ and $\sigma^2$ with at least one of the scale factors vanishing, at a finite value of the affine parameter. No known classical singularities of Bianchi-I model satisfy the conditions for this isolated case, and at this stage, its existence  is only a potential possibility. %allowed

\subsection{Strength of singularities}
The strength of singularities can be determined using the necessary and sufficient conditions obtained by Tipler \cite{tipler} and Kr\'{o}lak \cite{krolak}. These conditions are used to classify the singularities as strong and weak types, which provide insights on the magnitude of tidal forces experienced by an in-falling detector (or an observer) towards the singularity \cite{clarke-krolak}. %A singularity is considered weak if
According to the Kr\'{o}lak's criteria, the conditions to determine the strength of the singularities are the following. If for a null geodesic, the integral over Ricci curvature components,
\be\label{ricci_int}
\int_0^\tau \, \d \tau' R_{\mu \nu} \, u^\mu u^\nu
\ee
or the integral over Weyl curvature components
\be\label{weyl_int}
\int_0^\tau \, \d \tau' \left( \int_0^{\tau''} \, \d \tau'' |C_{\alpha \beta \mu \nu} \, u^\beta u^\nu|\right)^2
\ee
diverges, then the singularity is considered to be strong. Else the singularity is weak. Tipler's conditions are similar to those of Kr\'{o}lak, but involve an additional integral over the affine parameter \cite{clarke-krolak}. As an illustration of when a singularity is strong or weak, if in a physical evolution  $R_{\alpha \beta} u^\alpha u^\beta$ is proportional to $1/\tau^m$ and
$C_{\alpha\beta\mu\nu} u^\beta u^\nu$ is proportional to $1/\tau^n$, then  a singularity is strong curvature type by Kr\'{o}lak's criteria if $m \geq 1$ or $n \geq 3/2$. Similarly, a singularity is strong following Tipler's criteria if $m$ or $n \geq 2$. Thus, a singularity can be strong by Kr\'{o}lak's criteria, yet it can be weak according to Tipler's criteria. However, all singularities which are strong by Tipler's criteria are also strong by Kr\'{o}lak's criteria. Due to this reason, we will consider Kr\'{o}lak's criteria to address the resolution of strong curvature singularities in our analysis. \\

To analyze the strength of potentially singular events in effective spacetime description of LQC, we find the expressions for the integrands in the integrals (\ref{ricci_int}) and (\ref{weyl_int}). For the metric (\ref{metric}), the
only non-vanishing components of Ricci and Weyl tensors are $R_{ii}$ ($i = 1..4$) and $C_{1212}, C_{1313}, C_{1414}, C_{2323}, C_{2424}, C_{3434}$ (and those obtained by symmetric and anti-symmetric transformations on the indices, such as $C_{2121}, C_{2112}$ etc.). Using eq.(\ref{geod2}) %(where $x' = u^x, y' = u^y$ and $z' = u^z$)
with the expression for the Ricci tensor components in terms of first and second order time derivatives of the directional scale factors $a_i$, one can obtain the integrand for (\ref{ricci_int}) as
\ba\label{integrand_r}
R_{ab} \, u^a u ^b &=& \nonumber \f{k_x^2}{a_1^2} \, \left(H_1 H_2 + H_1 H_3 - \f{\ddot a_2}{a_2} - \f{\ddot a_3}{a_3} \right) ~ + ~  \f{k_y^2}{a_2^2} \, \left(H_1 H_2 + H_2 H_3 - \f{\ddot a_1}{a_1} - \f{\ddot a_3}{a_3} \right) \\
&& + \f{k_z^2}{a_3^2} \, \left(H_3 H_1 + H_2 H_3 - \f{\ddot a_1}{a_1} - \f{\ddot a_2}{a_2} \right) ~,
\ea
where $k_i$ are constants obtained from the geodesic equation (\ref{geod2}).\\

A similar calculation for the integrand of (\ref{weyl_int}) yields,
\ba\label{integrand_w}
C_{abcd} \, u^a u^b &=& \nonumber \f{k_x^2}{6 a_1^2} \Bigg[ (1 + a_1^2) \left(H_1 H_2 + H_1 H_3 - 2 H_2 H_3 + \f{\ddot a_2}{a_2} + \f{\ddot a_3}{a_3} - 2 \f{\ddot a_1}{a_1} \right) + 3(a_2^2 - a_3^2) \left(\f{\ddot a_3}{a_3} \right.\\ && \nonumber  ~~  \left. - \f{\ddot a_2}{a_2} + H_1 H_2 - H_1 H_3 \right)\Bigg] \, + \, \f{k_y^2}{6 a_2^2} \, \Bigg[(1 + a_2^2) \left(H_2 H_3 + H_1 H_2 - 2 H_1 H_3 + \f{\ddot a_1}{a_1}\right. \\
&& \nonumber ~~ + \left.  \f{\ddot a_3}{a_3} - 2 \f{\ddot a_2}{a_2}  \right) + 3(a_3^2 - a_1^2) \left(H_2 H_3 - H_1 H_2 + \f{\ddot a_1}{a_1} - \f{\ddot a_3}{a_3} \right) \Bigg] \\ && ~~ \nonumber + \f{k_z^2}{a_3^2}
\Bigg[(1 + a_3^2) \left(H_2 H_3 + H_1 H_3 - 2 H_1 H_2 + \f{\ddot a_1}{a_1} + \f{\ddot a_2}{a_2} - 2 \f{\ddot a_3}{a_3} \right) \\ && ~~ \nonumber + 3(a_2^2 - a_1^2) \left(\f{\ddot a_1}{a_1} - \f{\ddot a_2}{a_2} + H_2 H_3 - H_1 H_3 \right)\Bigg] ~\\
&& + \nonumber \f{1}{3} \Bigg[\left(k_x k_y - k_z \left(\f{k_x^2}{a_1^2} + \f{k_y^2}{a_2^2} + \f{k_z^2}{a_3^2}\right)\right) \left(H_2 H_3 + H_3 H_1 - 2 H_1 H_2 + \f{\ddot a_1}{a_1} + \f{\ddot a_2}{a_2} - 2 \f{\ddot a_3}{a_3} \right) \\
&& ~~ \nonumber + \left(k_y k_z - k_x \left(\f{k_x^2}{a_1^2} + \f{k_y^2}{a_2^2} + \f{k_z^2}{a_3^2}\right)\right) \left(H_1 H_2 + H_3 H_1 - 2 H_2 H_3 - 2 \f{\ddot a_1}{a_1} + \f{\ddot a_2}{a_2} + \f{\ddot a_3}{a_3} \right) \\
&& ~~ \nonumber + \left(k_z k_x - k_y \left(\f{k_x^2}{a_1^2} + \f{k_y^2}{a_2^2} + \f{k_z^2}{a_3^2}\right)\right) \left(H_2 H_3 + H_1 H_2 - 2 H_3 H_1 +  \f{\ddot a_1}{a_1} - 2 \f{\ddot a_2}{a_2} + \f{\ddot a_3}{a_3} \right)\Bigg].\\
\ea

In order to determine whether a singularity is strong in effective spacetime description of LQC, we substitute (\ref{integrand_r}) and (\ref{integrand_w}) in
 (\ref{ricci_int}) and (\ref{weyl_int}) respectively. Both of the above integrands, involve quadratic terms in the directional Hubble rates, linear terms in $\ddot a_i/a_i$ and
terms with inverse scale factors. Expressing $\ddot a_i/a_i$ in terms of $H_i$ and $\dot H_i$, integration by parts of
(\ref{ricci_int}) and (\ref{weyl_int}) yields integrals with terms with directional Hubble rates and $1/a_i^2$. Since, directional
Hubble rates are bounded functions (\ref{hib1_max}), whether or not the integrals in Kr\'{o}lak's conditions diverge depend on the behavior of $1/a_i^2$ terms. In particular, if in an evolution $a_i \rightarrow 0$ as the singularity is approached in a finite value of affine parameter, then integrals (\ref{ricci_int}) and (\ref{weyl_int}) {\it can} diverge, leading to a strong singularity \`{a} la Kr\'{o}lak.\\

Let us determine the strength of those curvature invariant diverging events in LQC which occur due to divergence in pressure at finite non-vanishing value of directional scale factors (or finite volume). For such `pressure' singularities, divergence of curvature invariants appears only due to divergence in $\ddot a_i/a_i$ terms. Since these terms are integrated over affine parameter at least once in the Kr\'{o}lak's conditions (and at least twice in the Tipler's) conditions, they do not yield divergences of the integrals (\ref{ricci_int}) and (\ref{weyl_int}). Pressure singularities occurring at a finite value of scale factors, thus, turn out to be weak singularities in the effective spacetime description of Bianchi-I model in LQC. This conclusion supplements the
result obtained in the previous sub-section where we established that for such singularities, geodesic evolution does not break down. Thus, these singularities are harmless -- they are weak in strength and do not forbid extension of geodesics. We recall that the properties of pressure singularities occurring at finite scale factors are similar to the isotropic models in LQC, where as for Bianchi-I model, they turn  out to be weak and geodesically extendible \cite{ps09}.

The only case where strength of the singular events in LQC can be strong is when at least one of the scale factors vanishes at a finite value of $\rho$, $\theta$ and $\sigma^2$ in the physical evolution at a finite value of affine parameter. This is the identical condition for the isolated case of geodesic inextendibility for a mathematically allowed possibility in effective dynamics of LQC found in Sec. IVA. If such an event turns out to be physically realizable in effective dynamics of LQC, then one can explicitly compute the integrals (\ref{ricci_int}) and (\ref{weyl_int}) and determine whether it corresponds to a strong or a weak singularity. \\

In conclusion, we find that those events which lead to a divergence in curvature invariants in LQC are weak singularities if they occur due to divergence in pressure at non-vanishing scale factors. If the divergence in curvature invariants is associated  with vanishing of one
 more scale factors at a {\it finite} energy density and expansion and shear scalars at a finite value of affine parameter, then the singularity can be strong or weak. We emphasize that examples of events of latter type in GR or LQC are not known, and this case may turn out to be physically not realizable.  %

\subsection{Fate of geodesics and strong singularities: physical examples}

We now discuss specific examples  where the implications of modified loop quantum dynamics on singularity resolution can be manifestly seen. We start with the case of a perfect fluid with an the equation of state $w > -1$ and a vanishing anisotropic stress. Perfect fluids with different values of $w$ have been extensively studied in the Bianchi-I models in GR and in many situations analytical solutions are also known \cite{bianchi_doro,bianchi_thorne,ellis_1967,jacobs_1968,ellis_mc1,mccallum}. Singularities in the classical theory, which are of barrel, cigar, pancake and point like, are strong and geodesically inextendible events occurring at a vanishing volume.  We will show that in  LQC,  all these singularities are resolved. We then discuss aspects of singularity resolution for potential curvature
invariant divergent events in Bianchi-I model in LQC which can arise for more general fluids satisfying a generalized equation of state $P = P(\rho)$, which in principle allow singularities other than those found for perfect fluids in GR. These would be anisotropic generalization of big rip, big freeze and sudden singularities found in isotropic models \cite{ssd,bfs,sudden}. Since explicit examples of equations of state which lead to these singularities in Bianchi-I model are not yet known, our discussion will only point out the general features of resolution of such potential singularities.

\subsubsection{Perfect fluid with $w > -1$}
 A perfect fluid with a vanishing anisotropic stress and a constant equation of state $w = P/\rho$ which is greater than -1 constitutes a large type of matter models which have been studied in Bianchi-I spacetimes in GR, which include dust ($w = 0$), radiation ($ w= 1/3$) and stiff matter ($w = 1$) \cite{bianchi_doro,bianchi_thorne,ellis_1967,jacobs_1968,ellis_mc1,mccallum}. In the effective dynamics of Bianchi-I model in LQC, quantum geometric modifications do not influence the matter part of the Hamiltonian constraint, and the stress-energy tensor of the matter content satisfies the conservation law $T^\mu_{~\nu;\mu} = 0$, where
%is described by a stress-energy tensor given by
\be
T_{\mu \nu} = (\rho + P) v_\mu v_\nu + P g_{\mu \nu} ~.
\ee
%\ee
%where the pressure is related to the energy density by $P = w \rho$ with $w$ a finite constant. Using the
The conservation law leads to
\be\label{rho_cons}
\dot \rho + (H_1 + H_2 + H_3) (1 + w) \rho = 0 ~,
\ee
which on integration gives,
\be\label{rho_sol}
\rho = C (a_1 a_2 a_3)^{-(1 + w)} ~
\ee
where $C$ is a constant determined by the initial conditions.

Let us recall some features of the classical evolution for the perfect fluids. For $w > -1$, energy density diverges as the mean scale factor $a = (a_1 a_2 a_3)^{1/3}$ approaches zero in a finite proper time. At these events directional Hubble rates diverge, causing  expansion and the shear scalar to become infinite in GR. Due to these divergences, expressions for Ricci (\ref{ricci}), Kretschmann (\ref{kretschmann}) and the square of the Weyl curvature (\ref{weyl1}) reveal that these curvature invariants grow unbounded as the singularity in classical theory is reached. The singularity, depending on $w$ and initial anisotropies can be of the form of a barrel, cigar, pancake or a point \cite{bianchi_doro,bianchi_thorne,ellis_1967}. Also, integrals (\ref{ricci_int}, \ref{weyl_int}) to determine the strength of singularities diverge and the singularities turn out to be strong. From the geodesic equations (\ref{geod1}) and (\ref{geod2}), we find that at these events, geodesic evolution breaks down. Thus, in classical GR, perfect fluids with $w > -1$ generically lead to physical singularities in Bianchi-I model.\\

We now analyze the existence of above classical singularities in LQC. Since the conservation law $(\ref{rho_cons})$ is unmodified in LQC, the resulting proportionality of energy density with the scale factors, turns out to be identical as in GR. However, unlike GR, where $\rho$ can grow unboundedly as $V \rightarrow 0$ for $w > -1$, it is bounded by a universal value in LQC.  %The universal bound on the energy density guarantees that $\rho$ can not diverge
Using the upper bound on energy density (\ref{rho_max}), eq.(\ref{rho_sol}) yields,
\be
C (a_1 a_2 a_3)^{-(1 + w)} \leq \f{3}{8 \pi G \gamma^2 \lambda^2} ~.
\ee
This inequality implies that in a physical evolution, volume $V = a_1 a_2 a_3$ never becomes smaller than a minimum value given by,
\be\label{vmin}
V_{\mathrm{min}} = \left(\f{8 \pi G \gamma^2 \lambda^2 C}{3}\right)^{\f{1}{1 + w}} ~.
\ee
 Given that $P = w \rho$, the bound on energy density also leads to a bound on the pressure of the perfect fluid: $P_{\mathrm{max}} = w \rho_{\mathrm{max}} \approx 0.41 w \rho_{\mathrm{Pl}}$. Recall that in Sec. III we showed that the curvature invariants in LQC can diverge only if the physical volume vanishes or the pressure of the matter content diverges. Using above results in eqs.(\ref{r_ci}, \ref{k_ci}) and (\ref{w_ci}), we find that for a perfect fluid with a finite equation of state $w > -1$, Ricci and Kretschmann scalars and the square of the Weyl curvature are always bounded in LQC.

The existence of a minimum volume (\ref{vmin}) in LQC, leads to another important implication. Note that all the classical singularities for a perfect fluid with $w > -1$, whether they are barrel, cigar, pancake or point like, occur at a vanishing physical volume \cite{bianchi_doro,bianchi_thorne,ellis_1967,jacobs_1968,ellis_mc1,mccallum}. Irrespective of the geometry of the singularity, a non-vanishing physical volume implies that the values of the scale factors where these singularities occur in GR are excluded by the effective dynamics in LQC. Along with the bounds on the directional Hubble rates, $H_{i ~\mathrm{max}} = 3/(2 \gamma \lambda)$ obtained from (\ref{H_bianchi1}), this ensures that the geodesic equations (\ref{geod1}) \& (\ref{geod2}) remain well defined in LQC. Thus, {\it for a perfect fluid with $w > -1$ geodesic evolution never breaks down in LQC.} This result is in sharp contrast to the behavior of geodesics for perfect fluids with $w > -1$ in GR, where irrespective of the choice of $w$ or initial conditions, classical spacetime is geodesically incomplete.

Finally, let us consider the integrands
(\ref{integrand_r}) and (\ref{integrand_w}). Since the vanishing of the scale factors $a_i$ in the effective dynamics of LQC is ruled out for perfect fluid with $w > -1$ and the directional Hubble rates are bounded, these integrands can only diverge if $\ddot a_i/a_i$ become infinite. However,  eqs.(\ref{ddota1}) (and similar equations for $\ddot a_2/a_2$ and $\ddot a_3/a_3$) imply that for a non-vanishing volume and finite pressure, $\ddot a_i/a_i$ are always bounded in LQC. Therefore, the integrands (\ref{integrand_r}) and (\ref{integrand_w}) are always bounded. The integrals (\ref{ricci_int}) and (\ref{weyl_int}) over these integrands are thus finite for any finite range of integration. Hence Kr\'{o}lak and Tipler's conditions for the existence of strong singularities fail to be satisfied.  We thus conclude that for a perfect fluid with $w > -1$, geodesic evolution is well defined for all times in LQC and there is neither a divergence of curvature invariants nor any strong singularities in the effective spacetime.

\subsubsection{Potential exotic singularities}

 For a perfect fluid with $w > -1$, singularities in classical theory occur when physical volume vanish.  However, if we consider a fluid with a generalized equation of state, singularities at finite volume are potentially possible.\footnote{For examples of models, in isotropic spacetimes in GR, where such generalized equations of state have been extensively studied, we refer the reader to Ref. \cite{ssd,bfs,sudden,not,dabrowski,yuron}.}
These singularities come in variety of forms and have been classified depending on the properties of Hubble rate, $\ddot a/a$ and higher order time derivatives \cite{not,dabrowski,yuron}. Since the Kr\'{o}lak's and Tipler's  conditions require at least the second time derivative of scale factor to diverge in order for a singularity to be potentially strong, we will restrict our discussion to only such events, i.e. to big rip, big freeze and sudden singularities. In the following, we assume that generalization of these singularities
exist in the Bianchi-I spacetime in GR and discuss the effects of loop quantization on them. As in the previous section, we assume a fluid with a vanishing anisotropic stress.\\

\noindent
(i) Big rip singularities: In isotropic models in GR, this singularity occurs when the scale factor diverges with a divergence in energy density and Hubble rate. In LQC, for isotropic models, these singularities have been shown to be absent \cite{sst,brip_lqc,ps09,sv}. Its generalization in Bianchi-I spacetime would involve a divergence of at least one of the scale factors, directional Hubble rate and  energy density, causing $\theta$ and $\sigma^2$ to become infinite. Due to these divergences, integrals (\ref{ricci_int}) and (\ref{weyl_int}) would diverge, and these singularities will be strong curvature type in GR (\`{a} la Kr\'{o}lak's criteria). Analysis of geodesic equations  (\ref{geod1},\ref{geod2}) shows that they would break down in the classical theory, due to divergence in directional Hubble rates. On the other hand, in the effective dynamics of LQC, energy density and directional Hubble rates are bounded, and hence the scale factors at which these divergences occur will be excluded from the effective spacetime. Thus, big rip singularities of Bianchi-I models in GR will be absent in LQC. \\%

\noindent
(ii) Sudden singularities: These singularities occur at a finite value of scale factor, energy density and Hubble rate but a divergent pressure in isotropic models \cite{sudden}. In the anisotropic setting, a sudden singularity would occur under similar conditions: a divergence in pressure at a finite value of the mean scale factor and energy density. From eq.(\ref{ddota1}) we find that this divergence would cause $\ddot a_i/a_i$ to blow up, resulting in curvature invariants (\ref{r_ci}-\ref{w_ci}) to become infinite. Since these singularities would occur at finite $a_i > 0$ and with a divergence only in the pressure, analysis of Sec. IVB shows that these will be weak singularities, both in GR as well as LQC.  Analysis of geodesic equations (\ref{geod1}) and (\ref{geod2}) shows that as in the isotropic case, these events will be geodesically extendible.\\

\noindent
(iii) Big freeze singularities: These singularities share the properties of big rip and sudden singularities. In isotropic models, these occur at a finite scale factor but with a divergence in energy density and Hubble rate. They are strong and geodesically inextendible in GR. In LQC, such singularities have been shown to be forbidden in isotropic models \cite{ps09,sv}. Their fate in Bianchi-I models will be similar. Due to bounds on directional Hubble rates and expansion scalar (\ref{H_bianchi1}) and (\ref{thetab1}), effective dynamics would cause a turn-around of scale factors before the classical singularity could be reached, and these singularities would be avoided.

In summary, if we allow generalizations of big rip, sudden and big freeze singularities  in Bianchi-I spacetime in GR, then of these only
sudden singularities are the ones that are not excluded in the effective spacetime description of LQC. Though at these events, curvature invariants in effective spacetime description of LQC diverge, these turn be weak singularities beyond which geodesics can be extended.

\section{Summary}

It is widely believed that a viable theory of quantum gravity must address the way space-like singularities of the GR are overcome. An important issue is whether  such a theory resolves {\it all} space-like singularities, in particular those beyond which geodesics in classical theory can not be extended, and which are strong by strength.\footnote{Questions about the resolution of other singularities can also be raised in quantum gravity. As an example, it has been argued that a viable theory of quantum gravity should not resolve a certain class of time-like singularities \cite{horo_myers}.}   The goal of this work, is to take a first step in answering this question for the loop quantization of Bianchi-I model. In recent years, resolution of big bang/crunch singularities has been successfully demonstrated in LQC for different models, and detailed analytical, phenomenological and numerical aspects of the quantum theory of various cosmological spacetimes have been studied. An effective spacetime description of the underlying quantum theory has also been derived, using geometrical formulation of quantum mechanics \cite{aa_ts}.  It is based on an  effective Hamiltonian \cite{jw,vt,psvt}, whose resulting dynamics turns out to have an excellent agreement with the quantum evolution in isotropic and anisotropic models \cite{aps2,aps3,apsv,bp,ap,madrid_bianchi1}. Effective dynamics has been extensively used to extract various novel physical predictions in LQC (see Sec. V of Ref. \cite{as} for a review).
These developments in LQC, set the stage to investigate the issue of generic resolution of strong singularities in Bianchi-I models.

 To understand the general nature of singularity resolution, key properties of curvature invariants, geodesic equations and strength of any singular events need to be analyzed. These were investigated in this work, assuming the validity of effective spacetime description, and for matter with a vanishing anisotropic stress.  Using effective dynamics of Bianchi-I model, it has been earlier shown that for generic matter, energy density ($\rho$) and directional Hubble rates ($H_i$) are bounded by universal values.  The latter lead to bounds on expansion ($\theta$) and shear ($\sigma^2$) scalars of geodesic congruences in the effective spacetime \cite{csv,cs09,gs1}. These bounds are direct ramifications of the underlying quantum geometry in LQC. Our results on the properties of curvature invariants show that though quantum geometric effects bind them in many cases of physical interest, interestingly, they can in principle diverge. Thus, the bounds on $\rho$, $\theta$ and $\sigma^2$ do not guarantee the boundedness of curvature invariants.  We show that curvature invariants are bounded in Bianchi-I model in LQC  except if in the physical evolution, pressure becomes infinite and/or the volume vanishes. Since these conditions must be satisfied at a finite value of $\rho$, $\theta$ and $\sigma^2$, divergences of curvature invariants occur under very special conditions. In isotropic models, for a generalized equation of state, divergence in pressure at finite scale factor and energy density results in a sudden singularity which is weak and a geodesically extendible event in classical theory as well as LQC \cite{ps09}. Generalization of these singularities in Bianchi-I models, satisfy the conditions for divergence in curvature invariants in LQC. However, unlike pressure singularities, examples of potential singularities which may occur at vanishing volume with a finite energy density and expansion and shear scalars are not known.

Existence of events where curvature invariants can potentially diverge, does not necessarily imply existence of a strong singularity or geodesic inextendibility. We show that this turns out to be true in the Bianchi-I model in LQC.
Analysis of the (null) geodesic equations in the effective spacetime reveals that geodesic evolution remains well defined for events where curvature invariants diverge due to pressure becoming infinite at finite scale factors. Such singularities also turn out to be weak, and thus are  harmless. These curvature invariant divergent events, thus do not lead to physical singularities in LQC. Only when the divergence in curvature invariants occurs at a vanishing volume with a finite value of energy density, and expansion and shear scalars, at a finite value of affine parameter, can the geodesic equations  in effective spacetime of Bianchi-I model in LQC break down. None of the known classical singularities of Bianchi-I models, satisfies this condition. At this stage, this isolated case is only a potential possibility which may not be physically realized in the effective dynamics. Assuming that such a curvature invariant diverging event is allowed by physical solutions, the resulting singularity may be strong or weak, depending on the details of the dynamical evolution. To understand the existence of such a potential singularity,
it will be important to understand the validity of effective spacetime description when scale factors approach zero.  If effective dynamics gets additional corrections at small scale factors,    above potentially curvature invariant diverging event may be ruled out.\footnote{If the topology is considered compact, terms originating from inverse scale factor corrections in loop quantization (see Ref. \cite{as} for details), would also potentially play an important role in this regime.} It is important to note that such a singularity does not arise for matter which obeys positive energy conditions. Hence, it is also possible that such a potential singularity is realized for unphysical conditions for matter in the effective dynamics, which may be restricted by the quantum theory.

  These results stand in sharp distinction to the ones in the classical theory where Bianchi-I models generically lead to strong singularities and the spacetime is geodesically incomplete. To illustrate this, we consider a perfect fluid with equation of state $w > -1$. In classical theory, this choice leads to barrel, cigar, pancake and point singularities  which are strong and geodesically inextendible \cite{hawking-ellis,bianchi_doro,bianchi_thorne,ellis_1967,jacobs_1968,ellis_mc1,mccallum}. In contrast, we show that all these singularities are resolved in LQC. Curvature invariants turn out to be bounded and geodesic evolution never breaks down for perfect fluid with $w > -1$ in LQC. We also discussed the potential generalizations of big rip, sudden and big freeze singularities in Bianchi-I spacetime. Analysis of the general features of these singularities using effective dynamics of Bianchi-I model shows that big rip and big freeze singularities are resolved by quantum geometric effects, whereas sudden singularities turn out to be of weak strength. As discussed above, these will correspond to pressure singularities  in Bianchi-I LQC where curvature invariants diverge. However, geodesics are extendible beyond them.

 Analysis in this work generalizes the one for the spatially flat isotropic models in LQC performed earlier \cite{ps09}. There
  we found that all strong singularities were resolved and the effective spacetime turned out to be  geodesically complete. Analysis of the effective dynamics for $k = \pm 1$ model, suggests the same result for spatially curved models \cite{sv}. We have now shown that similar results appear in the presence of non-vanishing Weyl curvature. To understand the physical implications of our analysis in more detail, it is important to include the cosmological constant and matter with equation of state $w < -1$. It is also important to study the implications for the vacuum Bianchi-I model where the singularities result only from the divergence in Weyl curvature. Though these can be considered as a straightforward generalization of our analysis, given the rich structure of singularities in classical Bianchi-I model, important subtleties can not be ruled out. Further, our  analysis assumed matter with a vanishing anisotropic stress. In future work, this assumption will be relaxed. Though we do not expect a qualitative change of results, presence of matter with a non-vanishing anisotropic stress is expected to enrich the phenomenological implications.

 %Results obtained in this analysis strengthen the expectation made by the author in Ref. \cite{ps09}, pertaining to the  resolution of all strong singularities in different cosmological models in LQC and in general in LQG.
  The genericity of these results on singularity resolution, achieved earlier for isotropic models \cite{ps09,sv} and here for Bianchi-I model,  suggest  existence of a non-singularity theorem in quantum gravity analogous to the singularity theorems in classical theory. Though these works can regarded as a first steps towards realizing such a non-singularity theorem,
  in future works, several steps need to be carefully and systematically taken as we consider more complex models to understand the underlying conditions of such a theorem. Some insights on these conditions arise from the recent work on detailed  contrasts on physics of anisotropic models performed in Ref. \cite{gs1}. There it was found that inclusion of spatial curvature in anisotropic setting, as in the case of quantization of Bianchi-II and Bianchi-IX models done in Refs.\cite{awe3,we1}, reveals two interesting features. First of these is the non-trivial role of energy conditions, and second is the role of inverse triad (or volume) corrections (in spatially curved models) on the physics of singularity resolution. Both of these features are expected to play an important role in proving such a theorem. These observations also make our expectations on genericness of non-singularity results stronger. First, it is well understood in LQC (see for eg. Ref. \cite{ps05}), that inverse volume modifications for model with positive curvature help the singularity resolution.\footnote{In fact, it is possible to obtain bounce in $k=1$ model with inverse volume  modifications, even if holonomy modifications are not considered  (see for eg. \cite{ps-topo}).} Thus, we expect that with modifications coming from both the holonomies yielding trignometric functions in the Hamiltonian constraint and the inverse triad effects, singularity resolution results would strengthen. Second, even in GR, singularity theorems are proved by demanding that matter satisfies certain energy conditions. Energy conditions can weed out unphysical solutions, such as those which violate weak energy conditions. Thus, inclusion of  energy conditions which restrict unphysical matter is expected to bring us closer to prove a non-singularity theorem on similar lines as the singularity theorems in the classical theory. %We expect that future work on the genericity of singularity resolution in symmetry reduced models in LQC will confirm above noted features.

Next steps in the above direction to prove genericity of singularity resolution would require going beyond the effective spacetime description, and the inclusion of inhomogeneities. For the first step, one will require a derivation of these results in full loop quantization. Though techniques exist to address these issues at the quantum theory level for some matter, these would require extensions to arbitrary matter or with certain energy conditions (such as the weak energy condition) to prove genericity of singularity resolution. So far little work has been done on the inclusion of inhomogeneities in loop quantization of symmetric models, though useful insights have been gained by treating inhomogeneities \`a la Fock quantization methods \cite{madrid}. Such a hybrid method, suggests resolution of singularities in a qualitatively similar fashion as in isotropic and anisotropic spacetimes in LQC. In future work, it will be important to understand various aspects of singularity resolution in such hybrid methods for arbitrary matter, both at the effective spacetime and quantum theory levels. Perhaps the most challenging step would then be to understand the full loop quantization of inhomogeneous spacetimes. In this direction, important insights are expected to arise by carrying forward works attempting to link LQG with LQC \cite{engle,tim}, and also by understanding detailed physics of incorporation of inhomogenities in LQG using spinfoam methods \cite{carlo}. Given the nascent stage of results on inclusion of inhomogenities in LQC and relation with LQG, at this stage, one can only speculate about the possible roadblocks on the road to prove a non-singularity theorem. On general grounds, we expect one of these to be related with the difficulties in expressing holonomies as almost periodic functions of connection, which can make analysis of the physics of singularity resolution more involved.   Fortunately, quantization of Bianchi-II \cite{awe3} and Bianchi-IX spacetimes \cite{we1}, performed without the feature of almost periodicity, and the resulting analysis of new physics \cite{gs1}, provide important hints to proceed in such a situation. It is hoped that future work on symmetry reduced models and on inclusion of inhomogeneities, on the lines of the analysis performed here, will provide vital clues and a deeper understanding  on this fundamental issue in quantum gravity.

\acknowledgments
This work is supported by NSF grant PHY1068743. We thank Brajesh Gupt for useful comments.

%***********************************************************************

%Never the less,  {\it assuming} that such mathematical possibilities  are allowed by the effective dynamics in LQC,
 %Assuming such a mathemtically possibility is realized,  definitive conclusion about strength of such singular event requires detailed knowledge of the way individual scale factors vanish at finite energy density and the expansion and shear scalars.
%\bibliographystyle{hunsrt}
%\bibliography{bib}

\end{document}